\documentclass[11pt]{article}
\usepackage{epsfig,amssymb,graphicx,amsmath,amsthm,subcaption}
\usepackage{authblk}
\usepackage{array}

\newtheorem{lemma}{Lemma}
\newtheorem{theorem}{Theorem}

\begin{document}

\title{Dynamics of vaccination in a time-delayed epidemic model with awareness}
\author{G.O. Agaba, Y.N. Kyrychko, K.B. Blyuss\footnote{Corresponding author: K.Blyuss@sussex.ac.uk}
}
\affil{Department of Mathematics, School of Mathematical and Physical Sciences,
University of Sussex, Falmer, Brighton BN1 9QH, UK}

\maketitle

\begin{abstract}
This paper investigates the effects of vaccination on the dynamics of infectious disease, which is spreading in a population concurrently with awareness. The model considers contributions to the overall awareness from a global information campaign, direct contacts between unaware and aware individuals, and reported cases of infection. It is assumed that there is some time delay between individuals becoming aware and modifying their behaviour. Vaccination is administered to newborns, as well as to aware individuals, and it is further assumed that vaccine-induced immunity may wane with time. Feasibility and stability of the disease-free and endemic equilibria are studied analytically, and conditions for the Hopf bifurcation of the endemic steady state are found in terms of system parameters and the time delay. Analytical results are supported by numerical continuation of the Hopf bifurcation and numerical simulations of the model to illustrate different types of dynamical behaviour.
\end{abstract}

\section{Introduction}
Vaccines are known to be effective means of disease control and prevention \cite{Farrington03, Haber95, Keeling13, Shim12v}, having led to a complete eradication of smallpox \cite{Bazin00,Li09} and a substantial reduction in the cases of polio, measles, mumps, rubella. Latest WHO forecasts suggest expected eradication of measles and mumps in Europe in the next few years \cite{who14}. Depending on a particular disease and each individual vaccine, the vaccine-induced immunity may be life-long, or individuals may require subsequent vaccinations to improve their immunity status. In order to achieve maximum impact, every vaccination campaign should be accompanied by appropriate information campaigns that educate individuals about the need for vaccination to prevent the spread of infection and achieve the desired level of herd immunity \cite{Keeling13}. In some cases, negative press coverage has led to a reduction in vaccine uptake or even complete disruption of the vaccination campaign, as has been the case with HPV vaccine in Romania \cite{Penta14} and the MMR vaccine in the UK \cite{Choi08}. Furthermore, fears associated with possible side effects or incorrect perceptions about vaccine efficiency may also be detrimental to the vaccine uptake and subsequent success \cite{Arino04,Shim12jtb}.

A number of mathematical models have looked into the dynamics of vaccination \cite{Anderson82,Arino04,don02,Hethcote00,Keeling13,K-Z00,Schenzle84,Shulgin98} focusing on different types of vaccination schedules, various scenarios of vaccine uptake and efficiency, and the resulting control of epidemics. Some work has also been done on developing techniques for assessment and quantification of vaccine efficacy and efficiency \cite{Farrington03,Haber95,Shim12v}. More recently, attention has turned to vaccination models that include different types of population awareness \cite{Li09,Sharma15,Tchuenche11} and/or time delays due to either epidemiological properties of infection, such as latency or temporary immunity, or time delay in individuals' responses to available information about the disease \cite{Abta14,Laarabi15,Meng10,Sekiguchi11}. Liu et al. \cite{Liu15} have recently discussed an interesting notion of ``endemic" bubble in the context of delayed behavioural response during epidemics, which corresponds to existence of periodic oscillations around the endemic steady state only for some finite range of basic reproduction numbers.

In this paper we focus on the interactions between two approaches to reducing population-level impact of an infectious disease: spread of awareness and vaccination. The literature on epidemic models of the concurrent spread of disease and information is quite substantial, and mostly consists of mean-field \cite{Agaba16,Funk09,Greenhalgh15,Kiss10,Manfredi13} or network models \cite{Funk10b,Funk09,Funk10,Gross08,Hatzopoulos,Juher15,Sahneh11,Wang13,Wu12}. Within the set of mean-field models, disease awareness can be treated as an additional ``media" variable \cite{Misra11,Misra15,Samanta13} or incorporated into reduced rates of disease transmission \cite{Cui07,Cui08,Li09,Liu07,Sun11,Tchuenche12,Tchuenche11}. Since there are several distinct contribution to the overall disease awareness that come from contacts between unaware and aware individuals, global awareness campaigns, or reports of the incidence of infection, it is often realistic to include in the models time delays that are associated with either delayed reporting of infected cases or delayed responses of individuals to available information about the disease \cite{Greenhalgh15,Misra11jbs,Zhao14,Zuo14,Zuo15}.

Zhao et al. \cite{Zhao14} have modelled the delay in media coverage of an epidemic outbreak as a delayed term acting to reduce the disease transmission rate. Greenhalgh et al. \cite{Greenhalgh15} have explicitly incorporated in their model a separate compartment for a level of disease awareness, and considered the effects of two time delays on epidemic dynamics, one associated with the ``forgetting time", i.e. the time it takes for the aware susceptible individuals to become unaware again, and the other being the time it takes for awareness to emerge from the infected cases being reported. Similar approach has been pursued by Zuo et al. \cite{Zuo14,Zuo15} who included the time delay in reporting of cases either through a delayed awareness term \cite{Zuo14}, or a delayed contribution from the infected cases to the growth of population awareness \cite{Zuo15}. Very recently, Agaba et al. \cite{Agaba17ec} have proposed and studied a model, in which population awareness increases due to global campaigns, as well as reported cases of infection, and a contribution from aware susceptible individuals. In this model, the time delay associated with delayed response of individuals to available information can lead to a destabilisation of endemic steady state and subsequent onset of stable periodic oscillations. Whilst these models have provided insights into epidemic dynamics with account for disease awareness and time delays associated with reporting of cases or modifying the behaviour, they did not consider the effects of epidemic control or vaccination.

In terms of analysis of control of epidemics in time-delayed models, Meng et al. \cite{Meng10} and Sekiguchi and Ishiwata \cite{Sekiguchi11} have considered the influence of pulse vaccination on dynamics of SIR epidemic models with time delay representing disease incubation time. Abta et al. \cite{Abta14} have studied a similar problem from the control theory point of view, and showed how optimal control of epidemics can be achieved. The main emphasis of these models was on the effects of vaccination on epidemics, in which some aspects of the disease transmission were delayed, but they did not make any account for disease awareness.

In this paper, we consider the effects of vaccination in an epidemic model with awareness. Of particular interest is the interplay between parameters characterising the emergence of disease awareness, the time delay associated with individuals' response to available information about the disease, and the levels of vaccination. The paper is organised as follows. In the next Section we introduce a time-delayed model of disease dynamics in the presence of awareness and vaccination, and establish the well-posedness of this model. Section 3 contains analytical results of feasibility and stability analyses of the disease-free and endemic equilibria, together with conditions for a Hopf bifurcation of the endemic steady state. Section 4 is devoted to a numerical bifurcation analysis and simulations that illustrate behaviour of the model in different dynamical regimes. The paper concludes in Section 5 with the discussion of results and open problems.

\section{Model derivation}

We begin by considering an SIRS-type model, which is a modification of the model analysed recently in Agaba \emph{et al.} \cite{Agaba17ec}. Unlike that earlier model, we now include vital dynamics (though the disease is still assumed to be non-lethal) and assume that vaccination may not confer a life-long immunity. The population is divided into groups of susceptible individuals unaware of infection, $S_n(t)$, susceptible individuals aware of infection, $S_a(t)$, infected individuals $I(t)$, and recovered individuals $R(t)$. There is a constant birth rate $b$, which is taken to be the same as the death rate, so that the total population $N$ remains constant, and it is assumed that all newborns are unaware and susceptible to infection. The disease is transmitted from infected to unaware susceptible individuals at a rate $\beta$, and this rate is reduced by a factor $0 < \sigma_s < 1$ for aware susceptibles, who take some measures to reduce their potential contact rate. Infected individuals recover at a rate $r$. Disease awareness $M(t)$ has contributions from the reported number of cases at a rate $\alpha_o$, from the aware individuals at a rate $\alpha$, and from some global awareness campaigns at a rate $\omega_o$, and the awareness is lost at a rate $\lambda_o$, whereas aware susceptibles lose their awareness at a rate $\lambda$. Finally, unaware susceptibles become aware at a rate $\eta$, and it is assumed that it takes time $\tau$ for them to become aware or to modify their behaviour in the relation to spreading infection. These assumptions lead to the following basic model
\begin{equation}\label{eqn1}
\begin{array}{l}
\displaystyle{S_n^\prime =  b N -\frac{\beta I S_n}{N} - \eta M(t - \tau) S_n + \lambda S_a - b S_n,}\\\\
\displaystyle{S_a^\prime =  -\frac{\sigma_s \beta I S_a}{N} + \eta M(t - \tau) S_n - \lambda S_a - b S_a,}\\\\
\displaystyle{I^\prime =  \frac{\beta I S_n}{N} + \frac{\sigma_s \beta I S_a}{N} - r I - b I,}\\\\
\displaystyle{R^\prime  =  r I - b R,}\\\\
\displaystyle{M^\prime  =  \omega_o + \alpha_o I + \frac{\alpha S_a}{N} - \lambda_o M,}
\end{array}
\end{equation}
where   $S_n(t)+S_a(t)+I(t)+R(t)=N={\rm const}$. A summary of model parameters is given in Table~\ref{Tab1}.

\begin{table}[!h]
\begin{tabular}{c|l}
Parameter & Definition\\
\hline
$b$ & birth rate (same as the death rate)\\
$\beta$ & disease transmission rate\\
$\sigma_s$ & rate of reduction in susceptibility to infection due to being aware\\
$r$ & recovery rate\\
$\alpha_o$ & growth rate of disease awareness from the reported number of infections\\
$\alpha$ & growth rate of disease awareness arising from aware individuals\\
$\omega_o$ & growth rate of disease awareness from global awareness campaigns\\
$\lambda_o$ & rate of loss of awareness generated by awareness dissemination\\
$\lambda$ & rate of loss of awareness in susceptible individuals
\end{tabular}\caption{Parameter definitions}\label{Tab1}
\end{table}

To investigate the effects of the introduction of a vaccine on the disease dynamics, we consider a situation where a proportion $v$ of newborns are vaccinated, and aware susceptible individuals are vaccinated at the rate $v_s$. With this assumption, $bvN\equiv v_i N$ newborns appear straight in the recovered (protected) class, and $b(1-v)N=(b-v_i)N$ newborns go to the class of unaware susceptibles. Note that the newborn vaccination is considered to be constant since infants, assumed to be unaware, are vaccinated as a result of awareness of their parents, as well as nurses/doctors. Once they become aware of infection as adults, that automatically classifies them as aware susceptibles that could then be influenced by the awareness campaign for enhancing vaccination courage. It is further assumed that after a period of time $1/\delta$, the individuals lose their immunity against the infection. If $\delta=0$, this describes a perfect vaccine, while $\delta>0$ describes a vaccine resulting in temporary (waning) immunity. Similar to some earlier works \cite{Agaba17ec, Zuo14, Zuo15}, it is assumed that upon losing immunity, a certain proportion, $p$, of individuals will join the aware susceptible class while the remaining proportion, $q=1-p$, will return to the unaware susceptible class. With these assumptions, a modified model has the form

\begin{equation}\label{eqn2}
\begin{array}{l}
\displaystyle{S_n^\prime= (b - v_i) N -\frac{\beta I S_n}{N} - \eta M(t - \tau) S_n + \lambda S_a + \delta q R - b S_n,}\\\\
\displaystyle{S_a^\prime=-\frac{\sigma_s \beta I S_a}{N} + \eta M(t - \tau) S_n - (\lambda + v_s + b) S_a + \delta p R,}\\\\
\displaystyle{I^\prime=\frac{\beta I S_n}{N} + \frac{\sigma_s \beta I S_a}{N} - (r + b)I,}\\\\
\displaystyle{R^\prime=r I + v_i N + v_s S_a - (\delta + b) R,}\\\\
\displaystyle{M^\prime=\omega_o + \alpha_o I + \frac{\alpha S_a}{N} - \lambda_o M.}
\end{array}
\end{equation}
This system has to be augmented by an appropriate initial conditions
\begin{equation}\label{eqn3}
\begin{array}{l}
S_n(0) = S_{n_0} \geq 0, \quad S_a(0) = S_{a_0} \geq 0, \quad I(0) = I_0 > 0, \quad R(0) = R_0 \geq 0,\\\\
M(0) = M_0 \geq 0,\qquad M(s) = M_0(s) \geq 0 \qquad \text{for all} \quad s\in [- \tau, 0).
\end{array}
\end{equation}
Before proceeding with the analysis of the model (\ref{eqn2}), it is essential to verify that it is feasible, i.e. its solutions remain non-negative and bounded for all $t\geq 0$.

\begin{theorem}
\label{thm1}
The solutions $(S_n(t), S_a(t), I(t), R(t), M(t))$ of the system (\ref{eqn2}) with the initial condition (\ref{eqn3}) are non-negative and bounded for all $t\geq 0$.
\end{theorem}

\noindent {\bf Proof.} Considering the equation for $I(t)$, let $t_i>0$ be the first time when $I(t)=0$, and the other components are still non-negative as per initial conditions, so
\[
S_n(t) \geq 0,\quad S_a(t)\geq 0\qquad \text{for all} \quad t \in [0, t_i].
\] 
Introducing an auxiliary quantity
\[
\psi = \min_{0 \leq t \leq t_i} \left\{ \frac{\beta S_n}{N} + \frac{\sigma_s \beta S_a}{N} - (r + b) \right\},
\]
we have the relation
\[
\displaystyle{I^\prime=\frac{\beta I S_n}{N} + \frac{\sigma_s \beta I S_a}{N} - (r + b)I\geq \psi I,}
\]
that can be readily solved to yield
\[
I(t_i)\geq I(0) e^{\psi t_i}>0,
\]
which gives a contradiction.

In a similar way, let us assume there exists a first time $t_n > 0$ such that $ S_n(t) > 0$ for $ t \in [0, t_n)$ and $S_n(t_n) = 0$, which implies $dS_n(t_n)/dt<0$. Substituting this value of $S_n$ into the first equation of the system (\ref{eqn2}) gives
\[
\frac{dS_n}{dt}\Big|_{t = t_n} = (b - v_i) N + \lambda S_a + \delta q R > 0,
\]
which contradicts the initial assumption. Consequently, $ S_n(t) > 0$ for $ t\geq 0$, and similar arguments can be used to establish that $S_a$, $R$ and $M$ remain non-negative for all $t\geq 0$.

Having established the positivity of all state variables, from the fact that $S_n(t)+S_a(t)+I(t)+R(t)=N={\rm const}$, it immediately follows that they are also all bounded between $0$ and $N$. Looking at the last equation of the system (\ref{eqn2}), we have
\[
M^\prime=\omega_o + \alpha_o I + \frac{\alpha S_a}{N} - \lambda_o M\leq \omega_o+\alpha_o N+\alpha-\lambda_o M,
\]
which can be solved to give
\[
\displaystyle{M(t)\leq M(0)e^{-\lambda_o t}+\frac{\omega_o+\alpha_o N+\alpha}{\lambda_o}\left(1-e^{-\lambda_o t}\right)\leq \widehat{M},}
\]
where
\begin{equation}\label{Mhdef}
\displaystyle{\widehat{M}=M(0)+\frac{\omega_o+\alpha_o N+\alpha}{\lambda_o}.}
\end{equation}
This suggests that throughout the time evolution, all solutions remain within the bounded region
\begin{eqnarray*}
\Phi  = \left\{(S_n, S_a, I, R, M)  \in \mathbb{R}_+^5: 0 \leq S_n, S_a, I, R \leq N, 0\leq M \leq \widehat{M} \right\}.
\end{eqnarray*}
\hfill$\blacksquare$

\section{Steady states and their stability}
The system (\ref{eqn2}) can have at most two steady states, a {\it disease-free equilibrium} and an {\it endemic equilibrium}. The disease-free steady state is given by
\[
E_0=(S_n^\circ, S_a^\circ, 0,R^\circ, M^\circ).
\]
where
\begin{equation}\label{eqn8}
S_n^\circ =  N h_n, \quad S_a^\circ = N h_a,\quad  R^\circ =N(1-h_a-h_n),\quad M^\circ = \frac{\omega_o + \alpha h_a}{\lambda_o},
\end{equation}
and
\[
\displaystyle{h_a = \frac{- x_2 + \sqrt{x_2^2 + 4 x_1 x_3}}{2 x_1},\quad h_n = \frac{\lambda_o [b - v_i + \delta q (1 - h_a) + \lambda h_a]}{\eta (\omega_o + \alpha h_a) + \lambda_o (\delta q + b)},}
\]
with
\[
\begin{array}{l}
x_1 = \eta \alpha (\delta + v_s + b),\\\\
x_2 = \lambda_o (\delta + b) (\lambda + b) + \lambda_o v_s (\delta q + b) + \eta \omega_o (\delta + v_s + b) - \eta \alpha (b + \delta - v_i),\\\\
x_3 = \lambda_o \delta p v_i + \eta \omega_o (b + \delta - v_i).
\end{array}
\]
The steady state $E_0$ is biologically feasible, as long as the following condition holds
\begin{equation}\label{ha_cond}
h_a < \frac{b + \delta - v_i}{\delta + v_s + b},
\end{equation}
which also implies $0<h_n<1$. Due to the fact that $h_a$ is monotonically increasing with $\omega_o$, we have
\[
0<h_a(\omega_o=0)<h_a<h_a(\omega_o=\infty)<\frac{b + \delta - v_i}{\delta + v_s + b},
\]
implying that a condition (\ref{ha_cond}) is always satisfied, so the disease-free steady state $E_0$ is biologically feasible for all values of parameters. Unlike $h_a$, $h_n$ is monotonically decreasing with $\omega_o$, and the same behaviour is exhibited by $h_a$ and $h_n$ in their dependence on $\alpha$. In fact, for large $\alpha$, $h_n\to 0$, while $h_a\to (b + \delta - v_i)/(\delta + v_s + b)$. An explanation for this is that increasing $\alpha$ leads to the growth of $M(t)$, which, in turn, results in the majority of susceptible individuals being aware, and they then contribute to further growth of $M(t)$ in a manner similar to the effect of a global awareness campaign.

The endemic equilibrium $E^*=(S_n^*, S_a^*, I^*, R^*, M^*)$ is given by
\begin{equation}\label{eqn13}
\begin{array}{l}
\displaystyle{S_n^* = N h_{n_*}, \quad S_a^* = N h_{a_*}, \quad I^* = N h_{i_*},}\\\\
\displaystyle{R^*=N(1-h_{n_*}-h_{a_*}-h_{i_*}),\quad M^* = \frac{N h_{i_*} \alpha_o + \omega_o  + \alpha h_{a_*}}{\lambda_o},}
\end{array}
\end{equation}
with
\[
h_{n_*} = \frac{y_2 \pm \sqrt{y_2^2 - 4 y_1 y_3}}{2 y_1}, \quad h_{a_*} = \frac{r + b - \beta h_{n_*}}{\sigma_s \beta}, \qquad h_{i_*}= \frac{h_{n_*} (m_1 h_{n_*} - m_6) + m_7}{\sigma_s \beta (\lambda_o \delta q + m_4 h_{n_*})},
\]
where
\[
\begin{array}{l}
m_1 = \beta \eta \alpha, \quad m_2 = \beta \lambda_o [\lambda + v_s + b + \delta p(1-\sigma_s)] + \eta \alpha (r + b) + \sigma_s \beta \eta \omega_o,\\\\
m_3 = \lambda_o [(\lambda + \delta p + v_s + b) (r + b) - \sigma_s \beta \delta p], \quad m_4 = N \eta \alpha_o + \beta \lambda_o,\\\\
m_5 =  \lambda_o (r + \delta p + b), \quad m_6 =\beta \lambda_o (\lambda + \sigma_s \delta q + \sigma_s b - \delta q) + \sigma_s \beta \eta \omega_o + \eta \alpha (r + b),\\\\
m_7 = \lambda_o [\sigma_s \beta (b + \delta q - v_i) + (\lambda - \delta q) (r + b)],
\end{array}
\]
and

\[
\begin{array}{l}
y_1 = m_1 (\lambda_o \delta q + m_5) + m_4 (m_6 - m_2), \quad y_2 = \lambda_o \delta q m_2 + m_5 m_6 + m_4 (m_7 - m_3),\\\\
y_3 = \lambda_o \delta q m_3 + m_5 m_7.
\end{array}
\]
The endemic steady state $E^*$ is biologically feasible, provided $0<h_{n_*}<(r + b)/\beta$.

Since the total population is constant, one can remove the equation for $R(t)$ and focus on the behaviour of variables $S_n(t)$, $S_a(t)$ and $I(t)$ only, with $R(t)=N-S_n(t)-S_a(t)-I(t)$. This reduces the total number of equations without affecting the system dynamics. We begin stability analysis of these steady states by looking at the disease-free equilibrium.

\begin{theorem}
\label{thm2}
The disease-free equilibrium of the system (\ref{eqn3}) is linearly asymptotically stable for all $ \tau \geq 0$ if the basic reproductive number satisfies the condition $R_v^d < 1$, where
\begin{equation}\label{RVDdef}
R_v^d = \frac{\beta (h_n + \sigma_s h_a)}{r + b}.
\end{equation}
\end{theorem}

\noindent {\bf Proof.} For $\tau=0$, application of the next generation matrix method \cite{DW} immediately yields the result of the Theorem.

For $\tau>0$, linearisation of the system (\ref{eqn3}) near its disease-free steady state gives the following characteristic equation for the eigenvalues $k$,
\begin{equation}\label{ceq_df}
(k + r + b - a_2 - a_5) [k^3 + x_4 k^2 + x_5 k + x_6 - a_0 (x_7 k + x_8)] = 0,
\end{equation}
where
\begin{equation}\label{agx_def}
\begin{array}{l}
\displaystyle{a_0 = e^{-k \tau}, \quad a_2 = \frac{\beta S^\circ_n}{N}, \quad a_4 = \eta S^\circ_n, \quad a_5 = \frac{\sigma_s \beta S^\circ_a}{N}, \quad a_6 = \eta M^\circ, \quad a_7 = \frac{\alpha}{N},}\\\\
g_1 = \lambda + b + a_6, \quad g_2 = \lambda_o + \delta + b, \quad g_3 = \delta + b + v_s, \quad g_4 = \delta q + b + a_6,\\\\
x_4 = g_1 + g_2 + v_s,\quad x_5 = g_1 g_2 + \lambda_o g_3 + v_s g_4,\quad x_6 = \lambda_o [v_s g_4 + g_1 (\delta + b)],\\\\
x_7 = a_4 a_7,\quad x_8 = a_4 a_7 (\delta + b).
\end{array}
\end{equation}

\noindent The first eigenvalue $k = a_2 + a_5 - (r + b)$ is negative whenever
\[
a_2 + a_5 - (r + b)<0\quad \Longleftrightarrow \quad \frac{\beta (h_n + \sigma_s h_a)}{r + b}  < 1\quad 
\Longleftrightarrow \quad R_v^d<1,
\]
with $R_v^d$ defined in (\ref{RVDdef}). Other eigenvalues can be found as the roots of the transcendental equation
\begin{equation}\label{cub_eq_exp}
k^3 + x_4 k^2 + x_5 k + x_6 = (x_7 k + x_8) e^{-k \tau}.
\end{equation}

Since the disease-free steady state is stable for $\tau=0$ whenever $R_v^d<1$, let us now investigate whether in this case stability can be lost for $\tau > 0$. To this end, we look for solutions of the equation (\ref{cub_eq_exp}) in the form $k = i \mu$. Separating real and imaginary parts gives
\begin{equation}\label{eqn17}
\begin{array}{l}
- x_4 \mu^2 + x_6= x_7 \mu \sin(\mu \tau) + x_8 \cos(\mu \tau),\\\\
- \mu^3 + x_5 \mu= x_7 \mu \cos(\mu \tau) - x_8 \sin(\mu \tau).
\end{array}
\end{equation}
Squaring and adding these two equations yields
\begin{equation}\label{zeq}
z^3 + y_4 z^2 + y_5 z + y_6 = 0,\qquad z=\mu^2,
\end{equation}
so if one can show that there are no real positive roots $z$ of this equation, then no eigenvalues of the equation (\ref{cub_eq_exp}) can even cross the imaginary axis, thus implying the stability of the disease-free steady state. We will once again use the Routh-Hurwitz criteria to show that all roots of the cubic equation (\ref{zeq}) have a negative real part, which is true if and only if
\begin{equation}\label{RHcon2}
y_4 > 0, \qquad y_5 > 0, \qquad y_6 > 0, \qquad y_4 y_5 > y_6.
\end{equation}
It is straightforward to show that the first three of these conditions holds,
\begin{align*}
y_4 & = x_4^2 - 2 x_5 = g_1^2 + \lambda_o^2 + (\delta + b)^2 + v_s^2 + 2 v_s (\lambda + \delta p + b) > 0,\\
y_5 & = x_5^2 - 2 x_4 x_6 - x_7^2 = (\lambda_o g_1 + a_4 a_7)(\lambda_o g_1 - a_4 a_7) + \lambda_o^2 [v_s^2 + (\delta + b)^2 \\
& \quad + 2 v_s (\lambda + \delta p + b)] + [v_s g_4 + g_1 (\delta + b)]^2  > 0,\\
y_6 & = x_6^2 - x_8^2 = (x_6 + x_8)(x_6 - x_8)=(x_6 + x_8)\left[\lambda_o v_s g_4 + (\delta + b) (\lambda_o g_1 - a_4 a_7)\right]> 0,
\end{align*}
and the fourth can be transformed into
\begin{align*}
y_4 y_5-y_6 & = (x_4^2 - 2 x_5) (x_5^2 - 2 x_4 x_6 - x_7^2) - x_6^2 + x_8^2\\
& = \Big[g_1^2 + (\delta + b)^2 + v_s^2 + 2 v_s (\lambda + \delta p + b) \Big] \Big[(\lambda_o g_1 + a_4 a_7)(\lambda_o g_1 - a_4 a_7)\\
& \quad +\lambda_o^2 \big[v_s^2 + (\delta + b)^2 + 2 v_s (\lambda + \delta p + b)\big] + \big[v_s g_4 + g_1 (\delta + b)\big]^2 \Big] + (a_4 a_7)^2\\
& \quad +\lambda_o^2 (\lambda_o g_1 + a_4 a_7)(\lambda_o g_1 - a_4 a_7) + \lambda_o^4 \big[v_s^2 + (\delta + b)^2 + 2 v_s (\lambda + \delta p + b) \big] > 0.
\end{align*}
which shows that $y_4 y_5 > y_6$, implying that all roots $z$ of the equation (\ref{zeq}) have a negative real part. Thus, there are no purely imaginary roots $k=i\mu$ of the characteristic equation (\ref{cub_eq_exp}), and the disease-free state $E_0$ is stable if $R_v^d < 1$ for any $\tau \geq 0$.\hfill$\blacksquare$\\

{\bf Remark.} {\it It is worth noting that stability of the disease-free steady state is not affected by the rate $\alpha_o$ of growth of awareness associated with the reported number of infections. The reason for this is that in the neighbourhood of the disease-free steady state, if $R_v^d<1$, the number of infected individuals would go to zero, thus reducing to zero its contribution to the growth of awareness, and therefore, it would have no further effect on the stability of $E_0$.}\\

Next, we turn our attention to the endemic steady state $E^*$. The characteristic equation for linearisation near this steady state has the form
\begin{equation} \label{eqn18}
k^4 + k^3 P_1 + k^2 (P_2 - a_0 x_7) + k (P_3 + a_0 \tilde{P}_3) + P_4 + a_0 \tilde{P}_4 = 0,
\end{equation}
with
\begin{align}\label{P14}
P_1 & = a_1 + a_3 + g_1 + g_2 + v_s,\nonumber\\
P_2 & = g_2 (a_1 + a_3 + g_1) +  v_s (a_1 + g_4) + \lambda_o g_3 + a_1 (\lambda + a_2 + a_3) + a_3 (a_5 + a_6),\nonumber\\
P_3 & = \lambda_o [v_s (a_1 + g_4) + (\delta + b)(a_1 + a_3 + g_1)] + a_1 [v_s (\delta q + a_2) +  r (\lambda + a_3 - \delta q)] \nonumber\\
& \quad + a_3 r ( a_6 - \delta p) +g_2 [a_1 (\lambda + a_2 + a_3) + a_3 (a_5 + a_6)],\nonumber\\
\tilde{P}_3 & = \alpha_o a_4 (a_1 - a_3) - x_7 (\delta + b + a_1),\nonumber\\
P_4 & =\lambda_o \big[a_1 [v_s (\delta q + a_2) +  \lambda (\delta + b) + r (\lambda + a_3 - \delta q)] + (\delta + b) [a_3 (b + a_1 + a_6) \nonumber\\
& \quad + a_2 (a_1 - a_3)] + a_3  r g_4\big],\nonumber\\
\tilde{P}_4 & = \alpha_o a_4 [v_s a_1 + (\delta + b) (a_1 - a_3)] -  x_7 a_1 (r + \delta + b),
\end{align}
and
\begin{equation}\label{agx_def_end}
\begin{array}{l}
\displaystyle{a_0 = e^{-k \tau}, \quad a_1 = \frac{\beta I^*}{N}, \quad a_2 = \frac{\beta S^*_n}{N}, \quad a_3 = \frac{\sigma_s \beta I^*}{N}, \quad a_4 = \eta S^*_n, \quad a_5 = \frac{\sigma_s \beta S^*_a}{N},} \\\\ \displaystyle{a_6 = \eta M^*, \quad a_7 = \frac{\alpha}{N},}  \quad g_1 = \lambda + b + a_6, \quad g_2 = \lambda_o + \delta + b,  \quad  g_3 = \delta + b + v_s, \\\\ g_4 = \delta q + b + a_6, \quad x_7 = a_4 a_7.
\end{array}
\end{equation}

For $\tau = 0$, the characteristic equation (\ref{eqn18}) turns into a quartic
\begin{equation}\label{eqn19}
\Rightarrow \qquad k^4 + k^3 P_1 + k^2 (P_2 - x_7) + k (P_3 + \tilde{P}_3) + P_4 + \tilde{P}_4 = 0.
\end{equation}
whose roots all have a negative real part if and only if the following Routh-Hurwitz conditions are satisfied
\begin{equation}\label{RHcon3}
\begin{array}{l}
P_1 > 0, \quad  P_2 - x_7 > 0, \quad P_3 + \tilde{P}_3 > 0, \quad P_4 + \tilde{P}_4 > 0,\\\\
P_1 [(P_2 - x_7) (P_3 + \tilde{P}_3) - P_1 (P_4 + \tilde{P}_4)] > (P_3 + \tilde{P}_3)^2.
\end{array}
\end{equation}
From the definitions of parameters in (\ref{P14}) it follows that
\[
P_1= a_1 + a_3 + g_1 + g_2 + v_s>0,
\]
and the condition $P_3 + \tilde{P}_3 > 0$ is always satisfied, provided
\begin{eqnarray}
\label{E_cond}
P_2 - x_7 > 0, \; P_4 + \tilde{P}_4 > 0, \; P_1 [(P_2 - x_7)(P_3 + \tilde{P}_3) - P_1 (P_4 + \tilde{P}_4)] > (P_3 + \tilde{P}_3)^2,
\end{eqnarray}
implying that all stability conditions for the endemic steady state with $\tau=0$ are given by (\ref{E_cond}). Before proceeding with verification of these conditions, one can note that 
\[
S^*_n > \frac{N \lambda_o (r + \delta p + b)}{N \eta \alpha_o + \beta \lambda_o} \quad \Longrightarrow \quad (N \eta \alpha_o + \beta \lambda_o) S^*_n > N \lambda_o (r + \delta p + b),
\]
which can be rewritten as
\begin{equation}\label{aaa}
\alpha_o a_4 + \lambda_o a_2 > \lambda_o (r + \delta p + b),
\end{equation}
and also
\begin{equation}\label{a1a3}
a_1 - a_3 = \frac{\beta I^*}{N} - \frac{\sigma_s \beta I^*}{N} = \beta h_{i_*} (1 - \sigma_s) > 0 \qquad \Longrightarrow \qquad a_1 > a_3.
\end{equation}
Using the relation
\[
(N \eta \alpha_o + \beta \lambda_o)S^*_n >N \lambda_o (r + \delta p + b),
\]
and the equations determining the components of the endemic steady state, one can find that
\begin{equation}\label{x7rel}
N \lambda_o (\lambda + v_s + b) S^*_a > \eta \alpha S^*_a S^*_n \qquad \Longrightarrow \qquad \lambda_o (\lambda + v_s + b) > \frac{\eta \alpha S^*_n}{N} =x_7,
\end{equation}
which can be used to verify the first stability condition in (\ref{E_cond}) as follows,
\begin{align*}
 P_2 - x_7 & = g_2 (a_1 + a_3 + g_1) +  v_s (a_1 + g_4) + \lambda_o g_3 + a_1 (\lambda + a_2 + a_3) + a_3 (a_5 + a_6) - x_7 \\
& >\lambda_o (\lambda + v_s + b) - x_7 > 0,
\end{align*}
which means that this condition is satisfied for any parameter values.

The second condition in (\ref{E_cond}) has the explicit form
\begin{align*}
P_4 + \tilde{P}_4 & =a_1 v_s [\alpha_o a_4 + \lambda_o a_2 - \lambda_o (r + \delta p + b)] + a_1 (r + \delta + b) [\lambda_o (\lambda + v_s + b) - x_7]\\
& \quad +\lambda_o a_1 r \delta p + (a_1 - a_3)(\delta + b)[\alpha_o a_4 + \lambda_o a_2 - \lambda_o (r + b)] + \lambda_o a_3 [r (\delta q + a_1)\\
& \quad + (\delta + b)(a_1 + a_6)] > 0,
\end{align*}
and again it is satisfied for any parameter values due to relations (\ref{aaa}), (\ref{a1a3}) and (\ref{x7rel}) shown above.

The last condition in (\ref{E_cond}) can be written as follows,
\begin{align}\label{P3P3}
(P_3 & + \tilde{P}_3) [P_1 (P_2 - x_7) - (P_3 + \tilde{P}_3)] - P_1^2 (P_4 + \tilde{P}_4) \nonumber\\
& = (P_3 + \tilde{P}_3) \big[(\lambda_o + v_s + a_3 + g_1) [\lambda_o (\lambda + v_s + b) - x_7] - \lambda_o [\lambda_o P_1 + (\lambda_o + v_s \nonumber\\
& \quad + a_3 + g_1)(\lambda + v_s + b)] + (a_1 + a_3 + g_1)[g_3 (\delta + b + a_1) + (\delta + b)(a_3 + g_1)] \nonumber\\
& \quad + a_3 [(a_5 + a_6)(v_s + a_3 + g_1) + r \delta p + \alpha_o a_4] + v_s (a_1 + g_4)(a_3 + g_1 + g_3) \nonumber\\
& \quad + a_1 [a_3 (a_3 + g_1) + r (\delta q + b) + g_1 (\lambda + b + a_3 - a_5) - \alpha_o a_4] + a_6 r (a_1 - a_3) \big] \nonumber\\
& \quad + P_1^2 \big[\lambda_o (\delta + b + a_1)[\lambda_o (\lambda + v_s + b) - x_7] + a_1 [\lambda_o^2(\delta + a_2) + x_7 (r + \delta + b)] \nonumber\\
& \quad + a_3 [(\delta + b)(\alpha_o a_4 + \lambda_o^2) + \lambda_o^2(a_1 + a_5 + a_6)] + \lambda_o^2(a_6 g_3 - v_s \delta p) \nonumber\\
& \quad + \alpha_o a_4 \lambda_o (a_1 - a_3)-\alpha_o a_1 a_4 g_3 \big] > 0.
\end{align}

Hence, we have the following result.

\begin{lemma}
Let the condition
\begin{equation}\label{end_t0_con}
(P_3 + \tilde{P}_3)  [P_1 (P_2 - x_7) - (P_3 + \tilde{P}_3)] - P_1^2 (P_4 + \tilde{P}_4)>0
\end{equation}
hold. Then the endemic steady state $E^*$ is linearly asymptotically stable for $\tau=0$.
\end{lemma}

\noindent {\bf Remark.} {\it Although it does not appear possible to prove that the condition (\ref{end_t0_con}) is automatically satisfied, extensive numerical simulations show that it does indeed hold for any parameter values, for which the endemic steady state $E^*$ is biologically feasible. Furthermore, numerical simulations suggest the endemic steady state $E^*$ is only biologically feasible, provided the condition $R_v^d>1$ holds.}\\

Having established stability of the endemic state $E^*$ for $\tau = 0$, the next step in the analysis is to investigate whether this steady state can lose stability for $\tau > 0$, in which case the characteristic equation (\ref{eqn18}) has the explicit form
\begin{equation}
\label{eqn20x}
k^4 + P_1 k^3 + P_2 k^2 + P_3 k + P_4 = (x_7 k^2 - \tilde{P}_3 k - \tilde{P}_4) e^{- k \tau}.
\end{equation}
In order for the steady state $E^*$ to lose its stability, some of the eigenvalues as determined by this equation must cross the imaginary axis. Looking for solutions in the form $ k = i \mu$ and separating real and imaginary parts gives
\begin{equation}\label{eqn20}
\begin{array}{l}
 \mu^4 - P_2 \mu^2 + P_4 = -(x_7 \mu^2 + \tilde{P}_4) \cos(\mu \tau)-\tilde{P}_3 \mu \sin(\mu \tau),\\\\
- P_1 \mu^3 +  P_3 \mu = (x_7 \mu^2 + \tilde{P}_4) \sin(\mu \tau) - \tilde{P}_3 \mu \cos(\mu \tau).
\end{array}
\end{equation}
Squaring and adding these equations yields the following equation for the Hopf frequency
\begin{equation}
f(\mu) = \mu^8 + y_7 \mu^6 + y_8 \mu^4 + y_9 \mu^2 + y_{10}=0,
\end{equation}
where
\[
y_7 = P_1^2 - 2 P_2,\hspace{0.1cm} y_8 = 2 P_4 + P_2^2 - 2  P_1 P_3 - x_7^2,\hspace{0.1cm} y_9 = P_3^2 - 2 P_2 P_4 - \tilde{P}_3^2 - 2 x_7\tilde{P}_4,\hspace{0.1cm} y_{10} = P_4^2 - \tilde{P}_4^2.
\]

Without loss of generality, one can assume that the equation $f(\mu)=0$ has eight different positive roots $\mu_j$, $j=1, \dots, 8$. For each of those roots, we can solve the system of equations (\ref{eqn20}) to find the corresponding value of the time delay $\tau$
\begin{equation}\label{eqn21}
\begin{array}{l}
\displaystyle{\tau_{n,j}= \frac{1}{\mu_j} \left[ \cos^{-1} \left(\frac{(P_2 \mu_j^2 - \mu_j^4 - P_4)(x_7 \mu_j^2 + \tilde{P}_4) + \tilde{P}_3 \mu_j^2 (P_1 \mu_j^2 -  P_3)}{(x_7 \mu_j^2 + \tilde{P}_4)^2 + \tilde{P}_3^2 \mu_j^2} \right) +  2 \pi n \right],}\\\\
j=1,\ldots,8,\qquad n = 0, 1, 2, \dots,
\end{array}
\end{equation}
and define
\begin{equation}\label{tau_mu0}
\displaystyle{\tau_0=\tau_{n_0,j_0}=\min_{1\leq j\leq 8, n\geq 1}\{\tau_{n,j}\},\quad \mu_0=\mu_{j_0}.}
\end{equation}
In order to establish whether the Hopf bifurcation actually occurs at  $\tau = \tau_0$, one has to determine the sign of $d\text{Re}[k(\tau_0)]/d \tau$. Differentiating the characteristic equation (\ref{eqn20x}) with respect to $\tau$ gives
\[
\left(\frac{d k}{d \tau}\right)^{-1} = \frac{(2 x_7 k - \tilde{P}_3) e^{- k \tau} - (4 k^3 + 3 P_1 k^2 + 2 P_2 k + P_3)}{(x_7 k^3 - \tilde{P}_3 k^2 - \tilde{P}_4 k) e^{- k \tau}} - \frac{\tau}{k}.
\]
Evaluating this at $\tau=\tau_0$ with $k=i\mu_0$,
\begin{figure}
\begin{center}
	\includegraphics[width = 14.4cm]{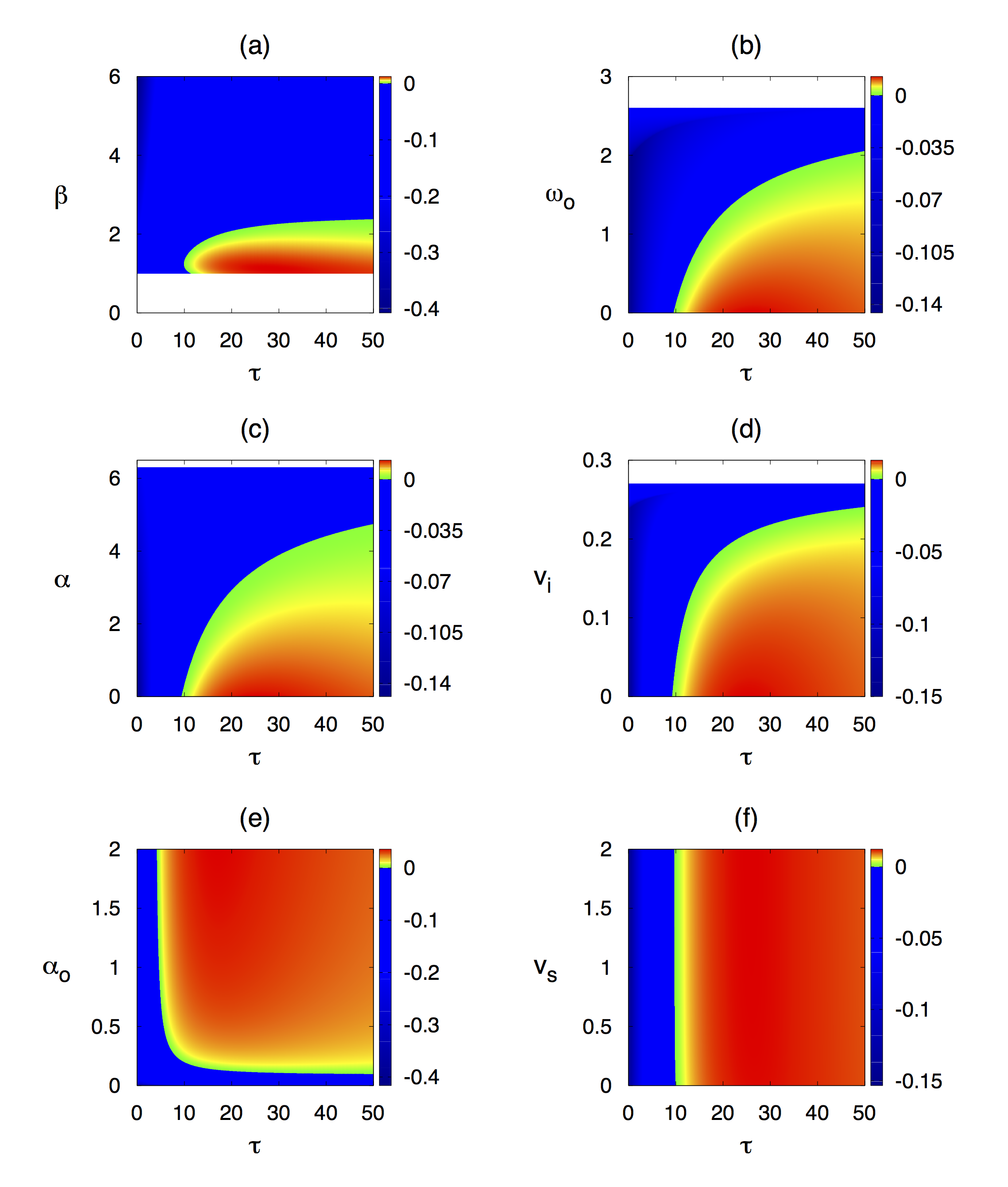}
	\vspace{-0.5cm}
	\caption{Stability of the endemic steady state $E^*$. Colour code denotes max[Re$(k)$], and in white regions the endemic steady state is not feasible. Baseline parameter values are as follows, $\beta = 1.2,\omega_o = 0.1,\alpha = 0.3,\alpha_o = 0.2,v_i = 0.04,v_s = 0.06$, other parameter values are $\lambda = 0.1, r = 0.2, \sigma_s = 0.04, p = 0.4, q = 0.6, \lambda_o = 0.3, \eta = 0.05, \delta = 0.3, b = 0.4, N = 100$.}\label{fig3}
\end{center}
\end{figure}
and using expressions for $\cos(\mu_0\tau_0)$ and $\sin(\mu_0\tau_0)$ in terms of coefficients $y_7$,..., $y_{10}$ of the characteristic equation (\ref{eqn20x}) yields
\[
\text{Re}\left(\frac{d k}{d \tau}\right)^{-1} \Big|_{\tau=\tau_0}= \frac{4 \mu_0^6 + 3 y_7 \mu_0^4 + 2 y_8 \mu_0^2 + y_9}{(x_7 \mu_0^2 + \tilde{P}_4)^2 + \tilde{P}_3^2 \mu_0^2}= z_v f^\prime(\mu_0),
\]
\[
z_v = \Big[2 \mu_0 \Big((x_7 \mu_0^2 + \tilde{P}_4)^2 + \tilde{P}_3^2 \mu_0^2 \Big)\Big]^{-1}.
\]
Since $z_v > 0$, this implies
\[
\text{sign} \left\{ \frac{d \text{Re}[k(\tau_0)]}{d \tau} \right \}= \text{sign} \left\{ \text{Re} \left(\frac{d k(\tau_0)}{d \tau} \right)^{-1} \right \}= \text{sign} [z_v f^\prime(\mu_0)]=\text{sign} [f^\prime(\mu_0)].
\]
This analysis can be summarised as follows.

\begin{theorem} \label{thm3}
Let the condition of {\bf Lemma 1} hold, and also let $\tau_0$ and $\mu_0$ be defined as in (\ref{tau_mu0}) with $f'(\mu_0)>0$. Then, the endemic steady state $E^*$ is linearly asymptotically stable for $\tau<\tau_0$, unstable for $\tau>\tau_0$ and undergoes a Hopf bifurcation at $\tau=\tau_0$.
\end{theorem}

\begin{figure}[!tb]
\begin{center}
	\includegraphics[width = 15cm]{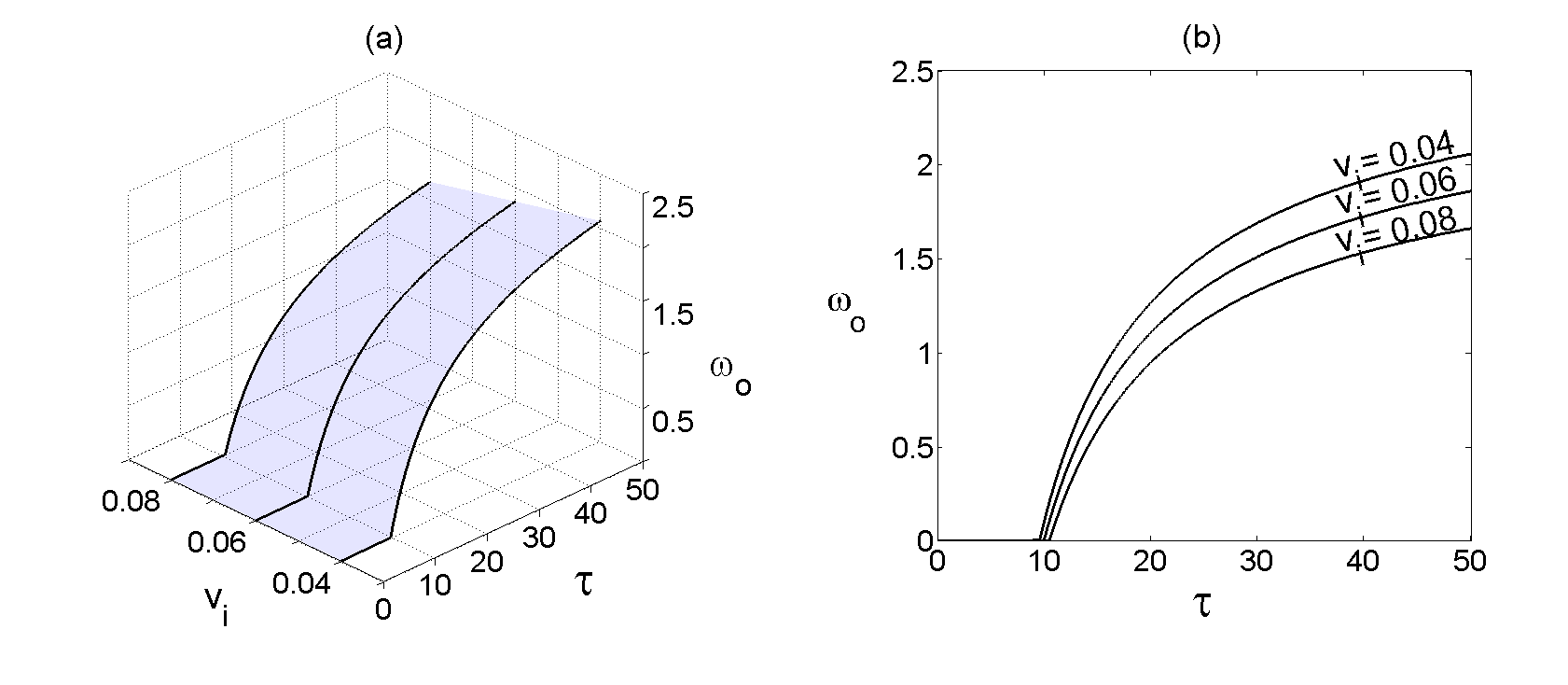}
	\vspace{-0.5cm}
	\caption{Stability boundaries of the endemic steady state $E^*$. The steady state is stable to the left of the surface in (a), and to the left of the lines in (b).  Parameter values are $\alpha = 0.3, \lambda = 0.1, \beta = 1.2, r = 0.2, \sigma_s = 0.04, p = 0.4, q = 0.6, \alpha_o = 0.2, \lambda_o = 0.3, \eta = 0.05, \delta = 0.3, v_s = 0.06, b = 0.4, N = 100$.}\label{fig5}\vspace{-0.7cm}
\end{center}
\end{figure}

\section{Numerical study of the model}

In order to better understand how different parameters affect the stability of the disease-free and endemic equilibria, we use a pseudospectral method \cite{Breda06} implemented in a traceDDE suite in MATLAB to numerically compute characteristic eigenvalues. Figure~\ref{fig3} illustrates how stability of the endemic steady state depends on the disease transmission rate $\beta$, local and global awareness rates $\alpha_o$, $\alpha$, $\omega_o$, and the time delay $\tau$ of individuals' response to available information. This figure shows that the endemic equilibrium only exists for a limited range of disease transmission rates, and it is stable for higher rates and unstable for smaller $\beta$. Increasing the awareness rate $\alpha_o$ leads to a destabilisation of the endemic steady state, but surprisingly, increasing a global awareness rate $\omega_o$ or a local awareness rates $\alpha$ actually results in stabilising an endemic steady state, whilst increasing these rates above certain values makes the endemic steady state unfeasible, in which case the disease-free steady state is stable. In terms of two types of vaccination, naturally, vaccination of aware individuals does not have any noticeable effect on stability of the endemic steady state, whereas increasing the vaccination rate of unaware individuals stabilises the endemic steady state, until it makes $E^*$ unfeasible and stabilises the disease-free steady state. Increasing the time delay $\tau$, in accordance with {\bf Theorem 3}, leads to de-stabilisation of the endemic steady state and the emergence of periodic solutions. Figure~\ref{fig5} further illustrates the stability boundary of the steady state $E^*$, showing that for higher vaccination rates, a lower rate of global awareness is required to stabilise the endemic steady state.

\begin{figure}[!tb]
	\includegraphics[width = 16cm]{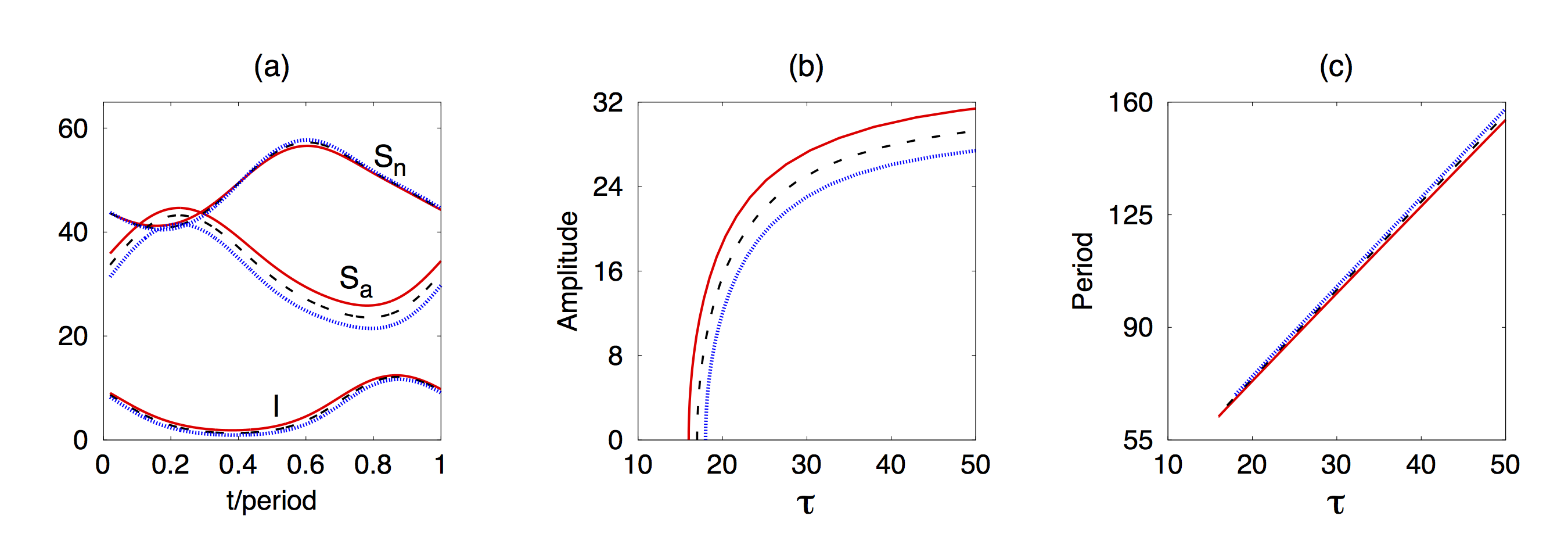}
	\vspace{-0.3cm}
	\caption{Bifurcation analysis of the endemic state: (a) periodic solutions showing the dynamics of $S_n$, $S_a$, $I$ variables only, (b) amplitude of periodic solutions depending on the time delay $\tau$, (c) period depending on time delay. In each plot, red solid lines correspond to $v_i = 0.04$, black dashed lines represent $v_i = 0.06$, and blue dotted lines correspond to $v_i = 0.08$. Other parameter values are $\alpha = 0.3, \lambda = 0.1, \beta = 1.2, r = 0.2, \sigma_s = 0.04, p = 0.4, q = 0.6, \omega_o = 0.1, \alpha_o = 0.2, \lambda_o = 0.3, \eta = 0.05, \delta = 0.3, v_s = 0.06, b = 0.4, N = 100$.}\label{fig10}
\end{figure}

Figure~\ref{fig10} demonstrates the results of numerical continuation of the Hopf bifurcation of the endemic steady state, as performed using DDE-BIFTOOL continuation software. It shows that both the amplitude, and the period of periodic solutions increase with the time delay $\tau$, and for higher vaccination rates $v_i$ the amplitude of the periodic solution is smaller, while the period is higher.

\begin{figure}[!tb]
	\includegraphics[width = 16cm]{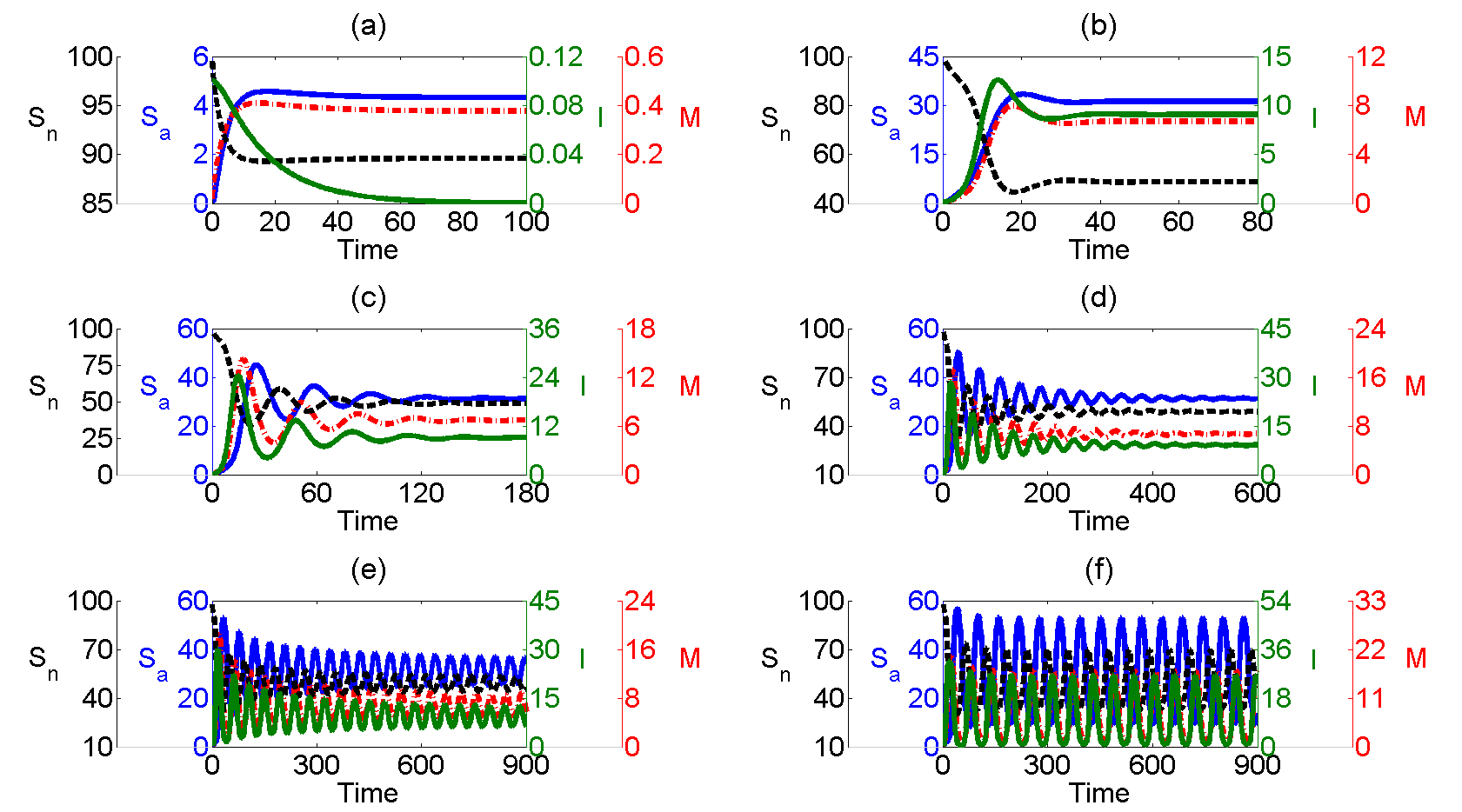}
	\caption{Numerical solutions of the system (\ref{eqn2}) (a) $\beta = 0.6, R_v^d = 0.8977, \tau = 0$, (b) - (f) $\beta = 1.2, R_v^d = 1.7955$. (b) $\tau = 0$, (c) $\tau = 4$, (d) $\tau = 8$, (e) $\tau = 10$, (f) $\tau = 16$. Other parameter values are $\alpha = 0.3, \lambda = 0.1, r = 0.2, \sigma_s = 0.04, p = 0.4, q = 0.6, \omega_o = 0.1, \alpha_o = 0.2, \lambda_o = 0.3, \eta = 0.05, \delta = 0.3, v_s = 0.06, v_i = 0.04, b = 0.4, N = 100$.}\label{fig1}
\end{figure}

In Fig.~\ref{fig1} we illustrate how actual dynamics of the system (\ref{eqn2}) changes depending on system parameters. Figure~\ref{fig1}(a) and (b) show the system approaches the stable disease-free or endemic steady states for $R_v^d < 1$ or $R_v^d >1$, respectively. One should note that according to {\bf Theorem 2}, the stability of the disease-fee steady state does not depend on the value of the time delay $\tau$, but rather on the basic reproduction number $R_v^d$ only, so if one keeps the value of $R_v^d<1$, the same kind of behaviour would be observed for any $\tau>0$. Choosing parameters in the range where $R_v^d >1$ and increasing the time delays $\tau$ results in the system approaching endemic steady state in an oscillatory manner, with the amplitude of oscillations increasing with the time delay. Once the time delay $\tau$ exceeds the critical value determined by {\bf Theorem 3}, the endemic steady state becomes unstable, and the system exhibits stable periodic solutions illustrated in Fig.~\ref{fig1}(f). The amplitude and period of such solutions themselves depend on the time delay, as has been shown earlier in Fig.~\ref{fig10}.

\section{Discussion}

This paper has analysed the effects of vaccination and different types of disease awareness on the dynamics of epidemic spread. We have studied analytically and numerically the conditions on system parameters which ensure feasibility and stability of the disease-free and endemic equilibria. These results suggest that stability of the disease-free steady state is independent of the time delay associated with the response of disease-unaware individuals to various types of awareness campaign, but it is rather determined by the basic reproduction number $R_v^d$ that depends on other epidemiological parameters, as well as awareness rates. On the contrary, stability of the endemic equilibrium does depend on the response time delay in such a way that while the endemic steady state is stable for $\tau=0$ (whenever it is biologically feasible), increasing the time delay can destabilise this endemic steady state and lead to the onset of stable periodic oscillations.

The numerical analysis has provided a number of insights into the relative roles of different parameters, some of which are natural, while others were surprising. Vaccination of aware individuals appears to not have a profound effect on the disease dynamics, while increasing the vaccination rates of unaware individuals (including newborn), can make the endemic steady state unfeasible, so that the disease would be eradicated, and the system would settle on a stable disease-free equilibrium. For large values of the time delay, reducing the rate of the disease transmission destabilises the endemic equilibrium, which should be expected. However, contrary to intuition, the same occurs when one reduces the rates of global awareness or local awareness, whereas one would expect that reduced awareness would support the maintenance of disease in the population, as is the case for the awareness stemming from the reported cases of disease. Moreover, increasing the rates of local and/or global awareness increases the time delay needed to destabilise the endemic steady state. Interestingly, all these different types of disease awareness only affect the stability of the endemic equilibrium for sufficiently large time delay, while for zero and small delays, the endemic steady state is always stable whenever it is feasible, regardless of the rates of awareness.

The results suggest that no matter how efficiently the cases of infection are reported, by itself this is not sufficient to create enough awareness to eradicate an epidemic, whereas global awareness campaign, and increasing the overall awareness level through contacts with other aware individuals are able to achieve this. Furthermore, the analysis shows an important role played by the vaccination of newborns, which can prevent epidemic outbreaks by providing a required level of herd immunity.

When assessing vaccine efficacy, one should be mindful of the fact that a vaccine may not provide complete protection against the disease. There are several approaches to modelling this, such as ``all-or-nothing" and ``leaky" vaccine scenarios \cite{Haber95,Halloran92,Shim12v}. ``All-or-nothing" vaccine is taken to represent a situation, where vaccine works only in some subset of vaccinated individuals, but for them it does provide complete protection. On the other hand, a ``leaky" vaccine describes a situation where all vaccinated individuals receive partial protection against the disease. In this paper, we have considered an idealised situation of a vaccine that takes on in all vaccinated individuals and provides complete protection but only for some period of time, i.e. a vaccine with waning immunity. Analysis of the effects of ``all-or-nothing" and ``leaky" vaccines will be the subject of further research.

\section*{Acknowledgements}
GOA acknowledges the support of the Benue State University through TETFund, Nigeria, and the School of Mathematical and Physical Sciences, University of Sussex.

\end{document}